\documentclass[letterpaper,twocolumn,10pt]{article}
\usepackage{usenix,epsfig,endnotes}
\usepackage{xcolor}
\usepackage{caption}
\usepackage{subcaption}
\PassOptionsToPackage{hyphens}{url}\usepackage{hyperref}
\usepackage{cleveref}

\begin{document}
\newcommand{\msnote}[1]{\textcolor{magenta}{Mridula: {#1}}}
\newcommand{\aanote}[1]{\textcolor{violet}{Aanjhan: {#1}}}
\newcommand{\hsnote}[1]{\textcolor{blue}{HS: {#1}}}
\newcommand{\mmnote}[1]{\textcolor{orange}{MM: {#1}}}

\newcommand{\eg}{e.g.,\xspace}
\newcommand{\bigeg}{E.g.,\xspace}
\newcommand{\etal}{\textit{et~al.\xspace}}
\newcommand{\etc}{etc.\@\xspace}
\newcommand{\ie}{i.e.,\xspace}
\newcommand{\bigie}{I.e.,\xspace}
\newcommand{\todo}[1]{{\color{red} #1}}
\newcommand{\idea}[1]{{\color{blue} #1}}
\newcommand{\captionfonts}{\small\bf}
\newcommand{\citneed}{[\textcolor{red}{cit}] }
\newcommand{\unit}[1]{\ensuremath{\, \mathsf{#1}}}
\newcommand{\ignore}[1]{}

\date{}

\title{\Large \bf Cryptography Is Not Enough: \\Real-time Location Spoofing of Authenticated GNSS Signals}

\author{Maryam Motallebighomi$^{*}$, Harshad Sathaye$^{*}$, Mridula Singh$\dag$, Aanjhan Ranganathan$^{*}$\\
$^{*}$Northeastern University, Boston, USA\\$\dag$ CISPA Helmholtz Center for Information Security, Saarbrücken, Germany}

\maketitle

\thispagestyle{empty}

\subsection*{Abstract}
In this work, we analyze the security guarantees of cryptographically protected GNSS signals and show the possibility of spoofing a receiver to an arbitrary location without breaking any cryptographic operation. 
Due to the increasing spoofing threats, Galileo and GPS are currently evaluating broadcast authentication techniques to validate the integrity of navigation messages. 
Prior work required an adversary to record the GNSS signals at the intended spoofed location and relay them to the victim receiver.  
Our attack demonstrates the ability of an adversary to receive signals close to the victim receiver and in real-time generate spoofing signals for an arbitrary location \emph{without modifying} the navigation message contents.
We exploit the essential \textit{common reception} and \textit{transmission time} method used to estimate pseudorange in GNSS receivers, thereby rendering any cryptographic authentication useless. 
We build a proof-of-concept real-time spoofer capable of receiving authenticated GNSS signals and generating spoofing signals for any arbitrary location and motion without requiring any high-speed communication networks or modifying the message contents.
Our evaluations show that it is possible to spoof a victim receiver to locations as far as 4000 km away from the actual location and with any dynamic motion path.
This work further highlights the fundamental limitations in securing a broadcast signaling-based localization system even if all communications are cryptographically protected. 

\section{Introduction}
Global Navigation Satellite Systems (GNSS) such as Galileo~\cite{galileo}, GPS~\cite{gps}, and GLONASS~\cite{glonass} are critical to a wide variety of applications ranging from navigation and tracking to modern communication and networking systems. 
It is well-known that civilian GNSS is vulnerable to signal spoofing attacks with increasing spoofing incidents observed in the wild~\cite{shanghai2019gpshack}.
In a GNSS spoofing attack, an adversary transmits radio-frequency signals that imitate legitimate satellite signals specifically crafted to force a receiver to compute a false location.
With the widespread availability of low-cost software-defined radio and public repositories~\cite{fernandez2011gnss}, the cost to spoof GPS signals has been significantly lowered (less than \$100).
Prior work has shown the possibility of changing the course of autonomous aerial~\cite{noh2019tractor}, terrestrial~\cite{tesla2019hack}, and aquatic~\cite{texas2013yacht} vehicles by simply spoofing GNSS signals.
Moreover, there are an increasing number of GPS signal interference and spoofing incidents~\cite{shanghai2019gpshack} being reported.
For example,  thousands of ships and GPS devices in Shanghai were suspected to be affected by GPS spoofing.  
It is also suspected that GPS spoofing resulted in several boats transmitting signals indicating they were sailing in circles off the California coast. In reality, they were thousands of miles away.

The lack of message authentication is a major contributing factor in generating fake satellite signals and falsifying a receiver's location.
To this extent, several countermeasures based on cryptographic authentication~\cite{scott2003anti} to protect against attackers generating spoofing signals are being proposed. 
For example, the recently launched Galileo's Open Service Navigation Message Authentication (OSNMA)~\cite{Guidelines_for_Test_Phase_v1.0}  authenticates the navigation message contents based on the TESLA protocol~\cite{perrig2002tesla} and one-way hash functions. 
As part of modernizing next-generation GPS, the United States Department of Defense is also exploring the use of Chips Message Robust Authentication (CHIMERA)~\cite{anderson2017chips}.
Both Galileo's OSNMA and CHIMERA digitally sign the navigation message contents and include the MAC within the message itself. 
In addition, CHIMERA \textit{replaces} parts of the spreading code with unknown bits called markers which are later revealed in a subsequent navigation message or an out-of-band channel. 
The above countermeasures aim to protect the integrity of the navigation message contents.
However, in GNSS, the user's location is computed based on both the navigation message contents \emph{and} its time of arrival.

This work analyzes the security guarantees of authenticated GNSS signals and shows that an attacker can spoof receivers to any location independent of the cryptographic primitive implemented.
Prior work~\cite{papadimitratos2008protection, psiaki2016gnss} showed the possibility of relaying GNSS signals (meaconing attacks) across large distances and spoofing the victim receiver's location to the location from where the legitimate signals were originally recorded.
In contrast, our attack does not require the attacker to capture legitimate signals at the location where the victim is to be spoofed.
We show how an adversary can spoof a victim receiver hundreds of kilometers away from its true location by temporally manipulating satellite signals received at the true location itself. 
Not only that, our attacker setup can, in real-time, generate and spoof dynamic motion paths independent of the location of the victim receiver. 
\emph{To the best of our knowledge, this is the first work that demonstrates the ability to spoof both arbitrary static and dynamic GNSS locations in real-time without modifying the contents of the navigation message, thereby rendering any cryptographic authentication useless.}
Our real-time setup takes less than 24 ms \footnote{20 ms to receive a bit, 317.485 $\mu$s of average processing time and 4 ms for sending the bits to attacker's transmitter} to generate spoofing signals from legitimate satellite signals, making the current delayed key-disclosure schemes incapable of detecting the attack.

Specifically, our attack works as follows.
First, we exploit the \emph{common reception time} and \emph{common transmission time} method that is fundamental to GNSS receiver designs for estimating the pseudoranges, i.e., the distance between a satellite and the receiver. 
The GNSS receivers typically assign a minimum travel time to the satellite signal arriving first (reference signal) and compute the pseudoranges based on the relative offsets of other signals to the reference signal. 
Our attack calculates the necessary delays to introduce in each satellite signal to achieve the required relative offsets corresponding to the specific spoofed target location.
This enables us to generate spoofing signals for \emph{any location or motion} using the signals received at the true location of the victim.
Second, it is necessary to generate the spoofing signals without decoding the entire satellite navigation message. 
Conventional receivers output navigation message contents every 6 s, making it harder to circumvent the time-binding of navigation message authentication primitives.
We designed our attacker to output the navigation message bits (note there is no encryption but only authentication) as it gets decoded every 20 ms.
Finally, the satellites are continuously in motion, and thereby the delays need to be constantly recomputed. 
Our attacker strategically chooses satellites to keep the re-computation time infrequent.

We designed and developed a real-time location spoofer using readily available software-defined radio platforms (less than \$1500). 
Our setup can receive legitimate GNSS signals and generate spoofing signals for any arbitrary location and motion in real-time in about $317.4~\mu s$ processing delay.
We sucessfully tested our attack on a commercial receiver (ublox M8N) and a software-defined GNSS receiver (GNSS-SDR) and show that it is possible to spoof a victim receiver to locations almost 4000 km away from the actual location without requiring any high-speed relay network or manipulating the message contents. We demonstrated our real-time setup in a video~\footnote{A video demonstration of this attack is available at \url{https://youtu.be/ylTpEsTCczs}}.
As a proof of concept for spoofing dynamic motion, we generate a 2.43 km dynamic motion path around a water reservoir 6 km away from the true location (also the location where the legitimate signals were recorded).
It is important to note that even though cryptographic signatures are transmitted as part of the navigation messages, as of today, commercial receivers lack the necessary infrastructure to validate and verify the transmitted signatures.
Hence, in our proof-of-concept attack, we show the feasibility of our proposed attack on conventional GNSS signals without manipulating the message contents.
We also evaluate the effect of the attacker's sampling rate, satellite constellation, and orbital motion on the accuracy and performance of the attack. 
Further, we show that with just two receivers (and network connectivity) strategically placed around the globe, an adversary can spoof a victim receiver to any location in the world.  
Thus, through this work, we further highlight the fundamental limitations in securing a broadcast signaling-based localization system even if all communications are cryptographically protected.  

\section{Background}
\subsection{GNSS Overview}
Global Navigation Satellite Systems (GNSS) is an umbrella term that refers to a satellite constellation providing positioning, navigation, and timing information to receivers on the ground.
USA's GPS, Europe's Galileo, Russia's GLONASS, and China's Beidou are some satellite navigation systems in operation today.
GNSS comprises a constellation of satellites equipped with high-precision atomic clocks that transmit ``navigation messages'' to the earth. 
Each satellite spreads the navigation messages using unique pseudorandom codes that are publicly available. 
The receiver receives these navigation messages and calculates the distance from the satellite to the receiver based on the transmission time contained within the navigation message and its time of arrival.
The receiver estimates its location using multilateration once it estimates its distances to at least four satellites.
The fundamental operating principle of all GNSS is the same, except they differ in the frequency of operation, precision, and availability of augmentation systems. 

\subsection{GNSS Spoofing Attacks}

A GNSS signal spoofing attack is a physical-layer attack in which an attacker transmits specially-crafted radio signals identical to legitimate satellite signals.
The goal of a signal spoofing attack is to force a victim receiver to compute a false location and/or time. 
GPS, Galileo, GLONASS, and Beidou are all vulnerable to spoofing attacks (as of today) due to the lack of signal authentication and publicly available pseudorandom codes, signal modulation schemes, and data-frame formats.
There are commercial signal generations~\cite{labsat} available today that can transmit multiple GNSS signals simultaneously.
Moreover, it is feasible to execute a signal spoofing attack with less than \$100 of hardware equipment due to the availability of low-cost software-defined radio platforms~\cite{ettus} and open-source GPS signal generation software~\cite{osqzss2015gpssim}.
The GNSS signal generators can transmit both static and entire trajectories, e.g., an adversary can spoof a stationary receiver to be in motion several kilometers away from its actual location.
In~\cite{tippenhauer2011requirements}, the authors present the fundamental requirements of executing a successful GPS spoofing attack.

\subsection{Cryptographic Countermeasures}
Several cryptographic countermeasures~\cite{wesson2012practical} were proposed as a means to prevent spoofing attacks. 
They can be broadly classified into two types: i) Navigation message authentication (NMA) and ii) Spreading Code Authentication (SCA). 
In NMA scheme, navigation messages are authenticated using digital signatures~\cite{margaria2017signal, curran2017message}.
and in SCA, random symbols (watermarks) are punctured in the public spreading sequence, which are later verified by the receiver. 
The increasing GNSS spoofing threat has forced GNSS operators to upgrade their existing infrastructure. 
For example, Europe's Galileo has started testing its open service navigation message authentication (OSNMA) and is available for public access~\cite{Guidelines_for_Test_Phase_v1.0} as of Jan 30, 2022.
Similarly, the US DoD is exploring the use of chips message robust authentication (Chimera) to improve the security of GPS, with several tests being conducted recently~\cite{anderson2017chips}.

\paragraph{Galileo OSNMA:}
Galileo's OSNMA is based on an adaption of the original timed efficient stream loss-tolerant authentication (TESLA) protocol~\cite{perrig2002tesla}.
Specifically, the navigation message is digitally signed, and the message authentication code is included using a set of \textit{40 reserved bits} of the navigation message. 
The key to verify the MAC is released after a delay, and the key itself can be verified using a previous key generated as part of a one-way chain defined in the TESLA protocol. 
The root key is kept secret, and therefore an adversary will not be able to generate the key-chain.
The delay ensures that the key used in the MAC generation procedure is not released until after the message and MAC are already received. 
Since OSNMA is based on a delayed key disclosure scheme, loose time synchronization at the receiver is critical and directly affects the scheme's effectiveness. 
To ensure the integrity of the navigation data offered by the TESLA protocol, OSNMA-enabled receivers verify the navigation data once the corresponding TESLA chain key is released by the satellite. 
This requires the receiver to be synchronized with a given accuracy to the Galileo system time.
According to OSNMA specifications \cite{fernandez2016galileospec, Guidelines_for_Test_Phase_v1.0, OSNMA_Info_Note}, the receiver is required to be i synchronization requirements at the receiver can range from 18 sec to 5 min.

\paragraph{GPS Chimera:}
Chimera includes NMA and SCA~\cite{anderson2017chips} modes of authentication.
Specifically, Chimera replaces some code chips with cryptographically generated markers and transmits the navigation messages.
The positions of the markers are also randomly chosen. 
The key used to generate the markers and their positions is revealed to the receiver after a certain time delay.
Additionally, the navigation message contents are also digitally signed, and the MAC is included as part of the message itself. 
GPS Chimera proposes two fundamental distribution mechanisms: i) Fast channel mode and ii) slow-channel mode.
The fast-channel mode uses an out-of-band high bandwidth network connection, while the slow-channel mode utilizes the navigation message to communicate the key.
The Chimera epoch for the fast-channel mode is the time duration over which a marker key is fixed.
The fast channel epoch is independent of the slow channel epoch and is $\approx$6 seconds~\cite{chimera_spec}.
In slow channel mode, the digital signature is included in an extra page, which makes the duration of the slow channel epoch around 3 minutes~\cite{chimera_spec}.
The US AirForce Research Laboratory is planning to launch Navigation Technology Satellite–3 (NTS-3) in 2023 to test Chimera ~\cite{hinks2021signal}. 
One of the modes of operation of GPS Chimera, which will be tested in this experiment, is ``NMA-only.'' The goal of these experiments is to test the effectiveness of data-only authentication. 
In addition, the NMA-only mode is designed to provide better tracking performance at the receivers~\cite{hinks2021signal}.

This work analyzes the security guarantees of the above-mentioned cryptographic measures to protect against spoofing attacks.
We note that these schemes are in various testing phases and are not yet available for broad public usage.
Through this work, we aim to raise awareness of the fundamental limitations of the proposed architectures and drive the research community to address these drawbacks in time for open public access.

\section{Spoofing Locations Using Authenticated GNSS Signals}

\begin{figure}
\centerline{\includegraphics[width=3.1in]{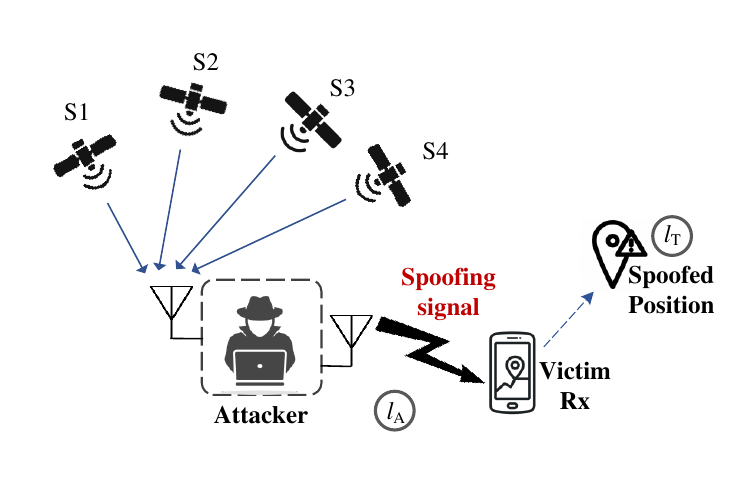}}
\caption{An overview of the proposed attack scenario. The attacker receives and re-transmits legitimate signals with strategically applied delays that force the victim receiver to compute a fake position.}
\label{attacker_model}
\end{figure}
\begin{figure*}[t]
\centering
\includegraphics[width=\linewidth]{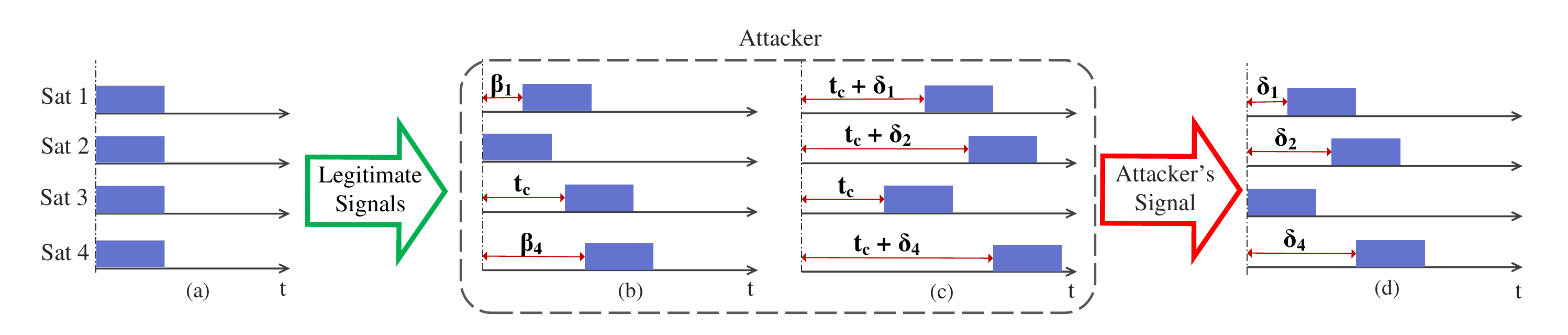}
\caption{A schematic describing the stages of the proposed attack based on manipulating the common reception time. (a) GNSS satellite transmission time (b) relative time of arrival in attacker receiver (c) relative times at attacker TX after attacker modifications (d) relative time of arrival in victim receiver.}
\label{common_reception}
\end{figure*}
\subsection{Attacker and System Assumptions}
The attacker's overall goal is to manipulate the estimated position at a victim receiver.
For example, the attacker is close to the victim receiver at location $l_A$ and intends to spoof the receiver to a target location $l_T$ as shown in Figure~\ref{attacker_model}.
We assume that the attacker has access to all public information, such as pseudorandom codes, signal modulation schemes, and data-frame formats. 
The attacker can receive legitimate GNSS signals, even if OSNMA and Chimera are enabled, as the spreading codes are public knowledge.
Additionally, the attacker can transmit the GNSS signal using the modulation and frame format expected at the victim receiver.
Finally, we assume that the attacker has sufficient transmission power to overshadow the legitimate signals\footnote{GPS's signal strength on the ground is typically -127.5 dBm}.
We emphasize that the attacker \emph{cannot} modify the contents of the navigation message as it would break the integrity, and the victim receiver will detect the manipulation.
Hence, in contrast to today's spoofing attacks, our adversary cannot generate navigation messages in advance.


We assume that the victim's receiver is a standard GNSS receiver, capable of decoding GNSS messages and validating the authenticity of the message content (e.g., Galileo OSNMA or GPS Chimera). 
The victim receiver can access confidential out-of-band information (e.g., Chimera's fast channel mode) and is loosely synchronized as required by the respective authentication scheme.
However, we assume that the victim receiver does not implement other non-cryptographic spoofing detection mechanisms, e.g., physical-layer-based spoofing detection techniques~\cite{ranganathan2016spree, warner2003gps, mcmilin2015gps}.
\subsection{Attack Overview}
Before describing the attack, it is crucial to understand how GNSS receivers determine their position by processing the satellite signals.
After pre-processing the received signal, the receiver first searches for visible satellite messages by correlating its own replica of the pseudorandom code corresponding to each satellite.
Once a satellite signal is detected, the receiver switches to tracking and demodulating the navigation message data for that specific satellite. 
The decoded data estimates the receiver's range or distance from each visible satellite. 
It is important to note that the satellite clocks are in tight synchronization while the receiver's clock (not using atomic clocks) contains errors and biases; therefore, we refer to the estimated ranges as pseudoranges. 
The receiver requires at least four pseudoranges to estimate its position.

To determine each pseudorange, the receiver needs the satellite signal's transmission and reception time.
The transmission time of each subframe is found in the navigational message. 
However, estimating the reception time of the signal~\cite{pini2012estimation} is a more involved process.
As shown in Figure~\ref{common_reception}a, signal transmission from the satellites is synchronized. 
Since the transmitted signals travel different distances, they arrive at the receiver with varying propagation delays (Figure~\ref{common_reception}b). 
Since the receiver does not have a high-accuracy reference clock as the satellites, the receiver uses the earliest arriving signal as the reference and computes the relative time difference of the remaining satellite signals. 
The result of this approach is not an absolute range for each satellite but a pseudorange relative to the first arriving reference satellite. 
Absolute pseudoranges are then estimated assuming a minimal travel time for the reference satellite based on known satellite orbits and typical user altitudes (e.g., for GPS, this is 65 to 85 ms).
Such an estimation of pseudoranges is fundamental to all GNSS receivers, and we exploit this design in our attack.

In a signal spoofing attack, an adversary can manipulate position estimation by either modifying the content of the navigation messages or the propagation delay. 
Since the message content is authenticated, we design our attack strategy to manipulate the reception time estimation method used for the pseudorange estimation. 
Suppose an adversary records and replays the GNSS signals as shown in prior work~\cite{shang2020flexible}, i.e., delays all the satellite signals by the same amount. In that case, the victim receiver's spoofed location is limited to where the adversary recorded the signal.

\begin{figure}[t]
\centering
\includegraphics[width=2.7in]{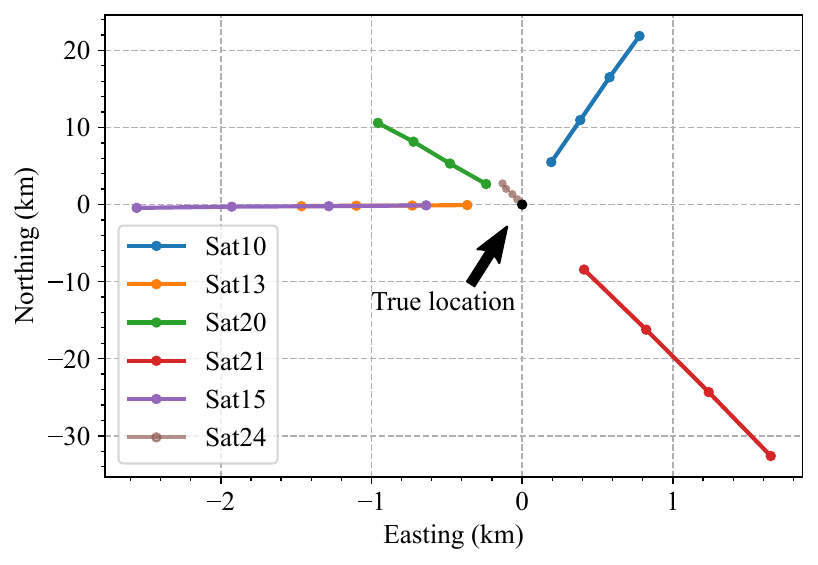}
\caption{The effect of delay applied to a single satellite. The magnitude and the direction of the shift depend on the delayed satellite's orientation relative to other non-delayed satellites. If we apply the same delay to all satellites, the obtained position will remain the same.}
\label{individual_sats}
\end{figure}

Figure~\ref{individual_sats} shows how delaying just one of the satellite signals moves the location estimated by the receiver, and delaying all the satellite signals does not cause any change to the actual location.
Therefore, in our attack, the adversary actively manipulates individual satellite signals' arrival time by introducing appropriate delays.
Given a set of satellite signals, our attacker continuously calculates and applies appropriate delays to spoof the victim to a specific location.
It is important to note that acquiring the legitimate signal and selecting delay values for each satellite signal is time-constrained when using OSNMA and Chimera. 
Recall that the victim considers the navigation message invalid and discards them once the keys for authentication are released. 
Furthermore, the satellites are in continuous motion. Therefore, the frequency of delay estimation directly impacts the satellite signals selected for temporal manipulation, and the achieved spoofed location accuracy.
The key modules that we design enable the attacker to overcome these challenges by providing access to navigation messages within the time constraints set by cryptographic countermeasures.
The following text describes our attacker modules in more detail.

\subsection{Key Components of the Attack}
\label{sec:components}
\begin{figure}[t]
\centering
\includegraphics[width=\columnwidth]{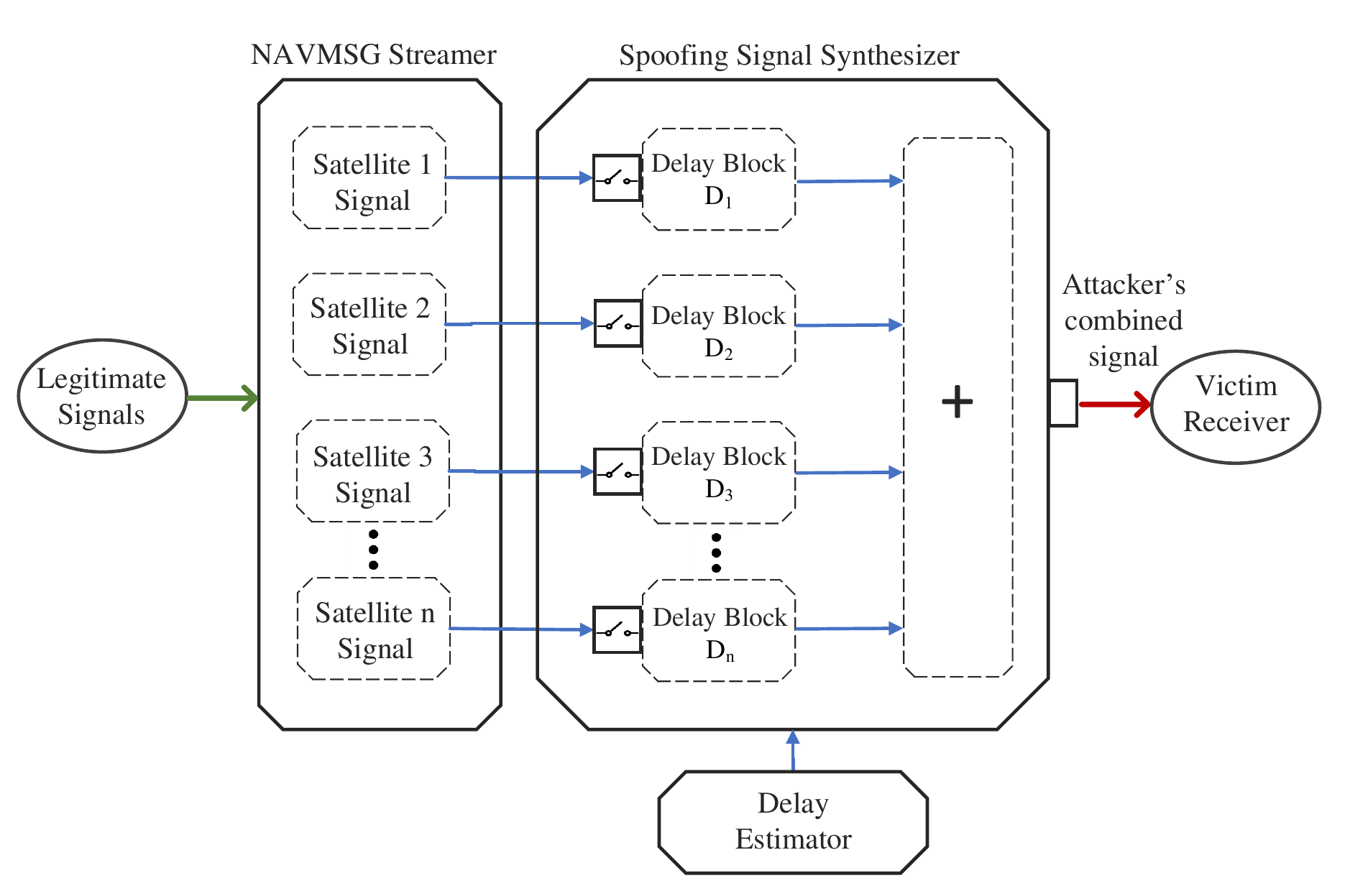}
\caption{A schematic representation of the entire signal processing pipeline. The NAVMSG streamer provides the required raw navigation bits, the delay estimator, and the signal synthesizer module that computes and applies the required delays.}
\label{attacker_modules}
\end{figure}

Our attack comprises of three key components as shown in Figure~\ref{attacker_modules}: i) NAVMSG streamer, ii) Delay Estimator, and iii) Spoofing Signal Synthesizer.
Recall that authentication mechanisms like OSNMA and Chimera enforce strict timing constraints, i.e., the receiver will discard navigation messages arriving after the disclosure of the key used to sign the message.
To achieve this, the NAVMSG streamer exploits the non-necessity of decoding the entire content of the navigation messages as the attack does not manipulate the navigation message data in any way. 
The delay estimator module calculates the necessary delays for each visible satellite signal to spoof the victim receiver to a target location.
The spoofing signal synthesizer module applies the delays computed by the delay estimator module to the satellite signals forwarded by the NAVMSG streamer. 
The spoofing signal synthesizer carefully selects satellite signals to apply the delays and combines them before spoofing the victim receiver during the synthesis process.

\paragraph{NAVMSG streamer.}
\label{sec:attack_navmsg_streamer}
The NAVMSG streamer is responsible for detecting visible satellite navigation messages and streaming them to the spoofing signal synthesizer in real-time.
In conventional receiver designs, the navigation message is output as a \emph{receiver observable} after the entire sub-frame is decoded, i.e., the signal has gone through the signal acquisition, demodulating, and decoding process, which takes 6\unit{s} for GPS and 30\unit{s} for Galileo~\cite{galileoNavMSG}. 
In our attack, it is necessary to detect and forward the navigation message signals as fast as possible for temporal manipulation. 
The goal is to hit the victim receiver with the spoofing signal \emph{before} the revelation of the appropriate authentication key.
We design the NAVMSG streamer to directly output the navigation message symbol from the receiver's tracking stage.
GNSS receivers perform correlation to identify visible satellite signals and synchronize before decoding the navigation message contents.
Our design uses the correlator output directly and streams the value as a single navigation message bit to the spoofed signal synthesizer.
This process eliminates the delays caused by other GNSS signal processing blocks.

With the method described above, the NAVMSG streamer can output a single navigation message bit every 20\unit{ms} for GPS and every 8\unit{ms} for Galileo.
GPS messages have a bitrate of 50 bps and 125 bps for Galileo.
At this rate, a receiver needs 20\unit{ms} and 8\unit{ms} to decode an individual navigation bit.
Also, the NAVMSG streamer separates each satellite signal using its unique pseudorandom spreading codes to allow the spoofing signal synthesizer to manipulate each satellite signal temporally.
For civilian-GNSS signals, the pseudorandom spreading codes are publicly known, allowing the possibility of acquiring each satellite's signals individually. 
If codes are not publicly available, signals originating from the different satellites can be separated using spatial methods like high gain antennas or antenna arrays ~\cite{zhang2018efficient,merwe2020gnss,akos2004high}. \\

\noindent \textbf{Delay Estimator.} The delay estimator calculates the delays to introduce in each satellite's signal received by the attacker at location $l_A$ such that when the spoofing signal is transmitted to the victim receiver, it computes the spoofed target location $l_T$.
To do that, we require the location coordinates  $\{x_i, y_i, z_i\}$ of a satellite $S_i$ at time $t$, which can be assumed to be public knowledge as it is part of the navigation message. 
Using this information, we can estimate the distance of location $l_A$ and $l_T$ from the satellite $S_i$ as: 

\begin{equation}
r_{A}^{i} = \sqrt{(x_{s}^{i} - x_{A})^{2} + (y_{s}^{i} - y_{A})^{2} + (z_{s}^{i} - z_{A})^{2}} 
\end{equation}
$$r_{T}^{i} = \sqrt{(x_{s}^{i} - x_{T})^{2} + (y_{s}^{i} - y_{T})^{2} + (z_{s}^{i} - z_{T})^{2}}$$
where $l_A(x_A, y_A, z_A)$ and $l_T(x_T, y_T, z_T)$ denote location coordinates for the attacker receiver and spoofed location. 
To spoof the location $l_T$, the attacker delays the signal originating from satellite $S_i$ and received at location $l_A$ by $\Delta\tau_{t}^{i}$.

\begin{equation}
 \Delta\tau_{t}^{i} = ( r_{T}^{i} - r_{A}^{i} - c*t_p )/ c
\end{equation}
where $t_p$ is the attacker's processing delay in receiving the legitimate signal, temporally manipulating, and transmitting it to the victim receiver.
We note that $t_p$ is constant for specific attacker hardware and is thus known to the attacker. 
We emphasize that the calculated delay is \emph{independent} of the distance between attacker and victim as these delays affect only the relative time offsets as indicated in Figure~\ref{common_reception}.
Since the satellites are in continuous motion, the attacker needs to continuously update the calculated delays, as it directly impacts the victim's obtained position and velocity. 
We present an evaluation of how the victim's estimated location is affected and its factors in Section~\ref{sec:eval}.\\

\noindent \textbf{Spoofing Signal Synthesizer.}
The spoofing signal synthesizer combines the individual satellite signals after applying the necessary delays computed by the delay estimator module. 
In other words, the synthesizer generates the spoofing signal to be transmitted to the victim receiver for the attack. 
A key function of the spoofing signal synthesizer is to comb through the available satellite signals and select the best satellites to include in the spoofing signal.
As we will show in Section~\ref{sec:eval_results}, the choice of satellites plays a significant role in the accuracy and performance of the spoofer. 
For instance, there are more than six visible GNSS satellites at any given time and location. 
However, using all the satellite signals will limit the maximum spoofable distance from the true location $l_A$.
It is also important to choose a subset of satellite signals that offer the \emph{lowest} geometric dilution of precision (GDOP).
GDOP is the geometry of the visible satellites in the sky and is low for a satellite constellation that is more spread out in the sky.
Remember that the adversary knows the true location $l_A$ and the spoofed target location ($l_T$) apriori and, therefore, the visible satellite signals to manipulate temporally.
Hence, the attacker can also compute the best subset of satellites with the highest GDOP in advance and use it during the attack.
The signal synthesizer also takes care of sanitizing the calculated delays. 
For instance, the attacker should avoid transmission of the signal where estimated delay $\Delta\tau_{t}^{i} < 0$.
If the calculated delay $\Delta\tau_{t}^{i}$ for some of the chosen satellites is negative,  the attacker picks the lowest negative delay value and adds it to all the other delays. The updated delay values will be as follows:
\begin{equation}
\hat{\Delta\tau}_{t}^{i} = \Delta\tau_{t}^{i} + t_{c}
\end{equation}
where $t_{c}$ is the common code phase offset and is equal to the lowest negative delay.
Also, choosing the satellites that are closer to the true location $l_A$ than to spoofed target location $l_T$ will result in smaller values for $t_c$.
We evaluate these parameters with real-world experiments in Section~\ref{sec:eval}.

\section{Attacker Implementation}
\label{sec:implementation}
\paragraph{NAVMSG Streamer.}
As described in Section~\ref{sec:components}, it is necessary to detect the presence of a satellite's navigation message well within the time constraints set by the message authentication scheme, which is 1.5 sec or 3 min for GPS Chimera~(based on the mode of operation) and 30 sec for Galileo.
Commercial GPS receivers like uBlox provide access to raw navigation messages. 
However, the user has to wait for 6 s for GPS and 30 s for Galileo for each navigation subframe, making it unsuitable for use for our attack.
In our work, we leveraged the design of GNSS-SDR and implemented the NAVMSG streamer as part of its telemetry decoder module.
The telemetry decoder module provides access to raw navigation message symbols directly at the correlator output after the receiver detects the presence of a specific satellite.
Additionally, we timestamp the navigation symbols and stream them to the spoofing signal synthesizer. 
Each NAVMSG streamer message includes the navigation message, our timestamp, and the corresponding satellite identification PRN.
Furthermore, the NAVMSG streamer separates the satellite signals before streaming them to the spoofing signal synthesizer. 
Here, we note that prior work has demonstrated the ability to separate satellite signals using directional antennas even if the pseudorandom codes are kept secret.

\paragraph{Delay Estimator and Spoofing Signal Synthesizer.}
The time delays to apply to each satellite signal are estimated based on the location where the attacker records the legitimate signals, the target location to spoof, and the satellite's orbital status.
The satellite orbital information is typically public knowledge or predicted based on the previously decoded navigation messages. 
The method for calculating the required delays is described in Section~\ref{sec:components}.

\begin{figure} [t]
\centerline{\includegraphics[width=3.2in]{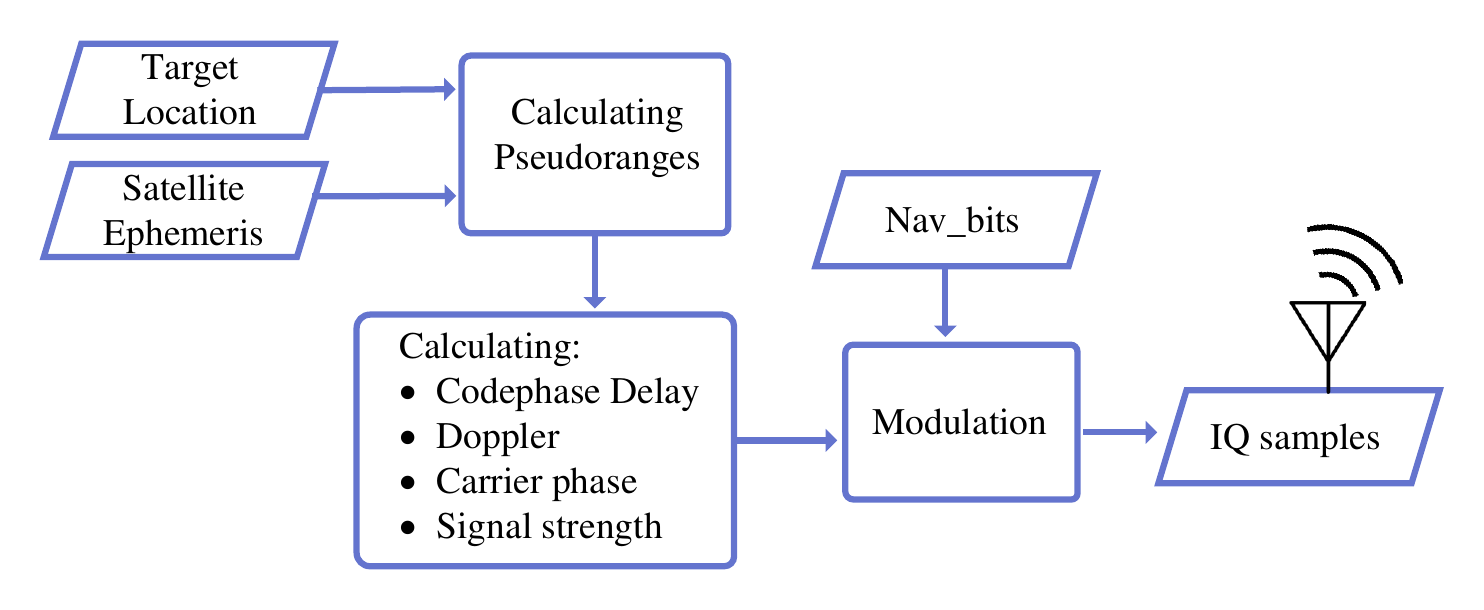}}
\caption{A flowgraph showing the implementation of our spoofing signal synthesizer. The desired pseudoranges and necessary signal parameters like carrier phase, code phase delay and frequency are calculated using satellite ephemeris and the target location. These calculated signal parameters are used to modulate the navigation bits received from the NAVMSG streamer and generate the spoofing signal.}
\label{using_delay_flowgraph}
\end{figure}

The spoofing signal synthesizer accepts a location, satellite ephemeris and received raw navigation bits and generates the required spoofing signal.
The architecture of the signal synthesizer is depicted in Figure~\ref{using_delay_flowgraph}.
Based on the provided location and current satellite positions we first calculate the ranges to all the visible satellites.
Next, the calculated range is used to obtain the code phase delay and the carrier phase measurements.
The necessary Doppler shift is calculated from rate of change of pseudorange and the wavelength of carrier frequency.
These calculated parameters along with the PRN code is used in modulation of the bits received from the NAVMSG streamer.
It is important to note that the signal synthesizer is required to perform all these calculations periodically to account for the satellite's motion over time

As shown in Figure~\ref{using_delay_flowgraph}, the spoofing signal synthesizer was configured to accept calculated delays at run-time, enabling complete control of the spoofing target location. 
We elaborate on the importance of real-time delay manipulation in Section~\ref{sec:attack_navmsg_streamer}. 
The synthesizer receives the satellite signals from the NAVMSG streamer and selects a subset of the satellites, as shown in Figure~\ref{attacker_modules}.
The estimated delays are applied to each satellite signal, combined, and transmitted to the victim receiver.

\section{Experimental Evaluation}

\label{sec:eval}
In this section, we evaluate the performance of our attack. First, we describe the evaluation setup and metrics used to measure our attack performance. Then, we provide an example spoofing scenario that we tested using the experimental setup for both GPS and Galileo signals. Finally, we discuss the results of our experiments.


\subsection{Experiment Setup}
We test our attack on a commercial receiver (ublox M8N) as well as GNSS-SDR~\cite{fernandez2011gnss}.
GNSS-SDR is an open-source software-defined GNSS receiver based on GNU Radio~\cite{gnuradio} capable of detecting, synchronizing, demodulating, and decoding the navigation messages originating from the constellations like GPS, Galileo, GLONASS, and BeiDou. 
It can process raw GNSS signals from a file source or from SDRs such as USRP~\cite{ettus} and enables us to gain deep insights into the attack performance.
\begin{figure}[t]
\centerline{\includegraphics[width=3.0in]{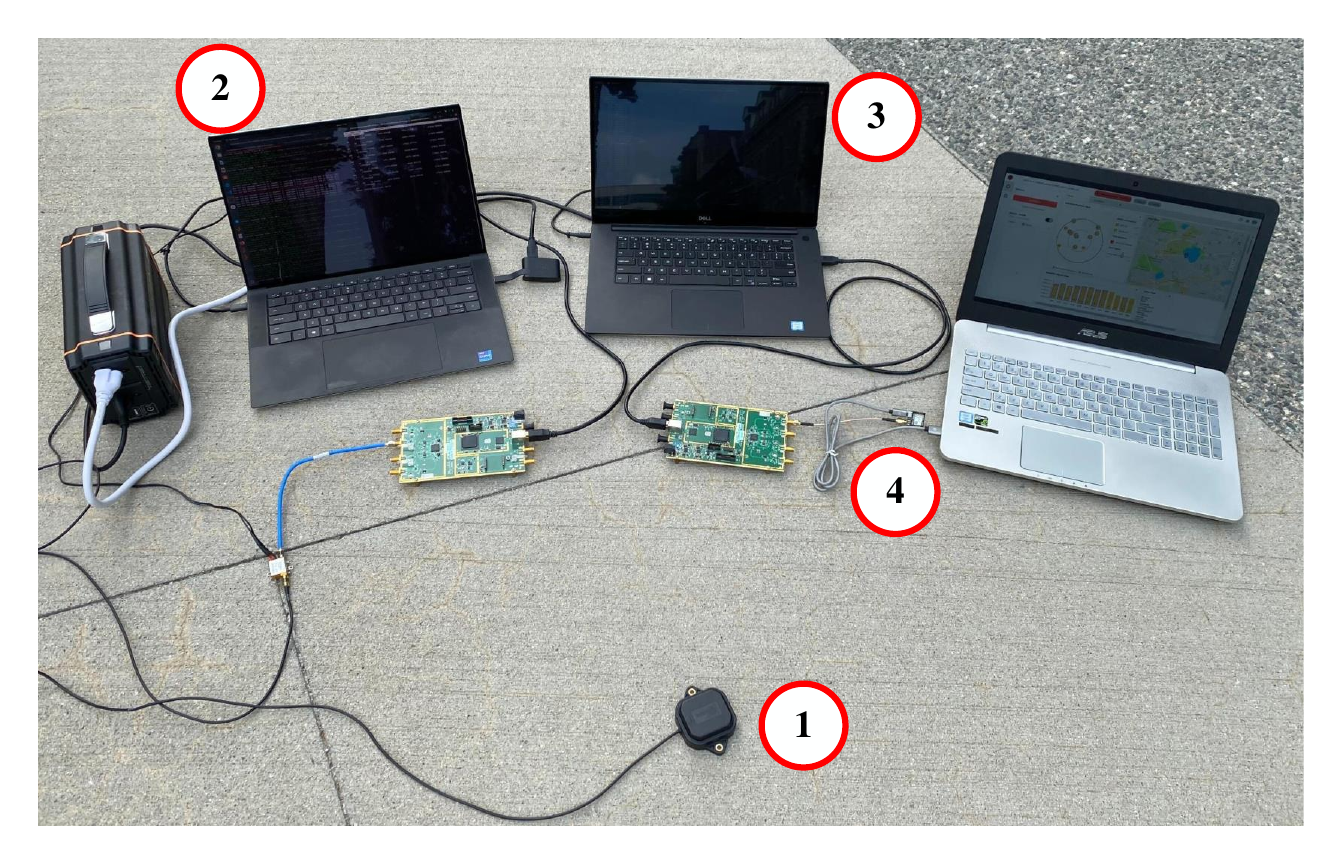}}
\caption{Experimental setup showcasing the real-time relay system: 1) Active GPS antenna with a 5V bias-tee, 2) NAVMSG streamer, 3) signal synthesizer and 4) ublox M8N GNSS receiver}
\label{real-time-setup}
\end{figure}

Figure~\ref{real-time-setup} shows our setup that is capable of manipulating live over-the-air GPS signals~\footnote{A video demonstration of this attack is available at \url{https://youtu.be/ylTpEsTCczs}}.
An active GNSS antenna feeds live GPS signals to GNSS-SDR that uses a USRP B210 as its RF front-end.
A streamer client connects to GNSS-SDR and streams the decoded navigation bits to another laptop that runs the signal synthesizer.
Based on the spoofed location, the signal synthesizer modulates the received bits as described in~\Cref{sec:implementation}.
It interfaces with another USRP B210 for transmission of the generated signal.
Finally, the signal is fed to a uBlox GPS receiver.
In addition to live over-the-air signals, we also use a real-time GPS signal generator that generates a continuous stream of IQ samples that are transmitted using another SDR.
Replacing the GNSS antenna with a signal generator ensures repeatable simulation conditions, provides complete control over the signal properties, and enabled us to investigate the effect of different parameters on the effectiveness of our attack.

\begin{figure}[t]
\centerline{\includegraphics[width=3.1in]{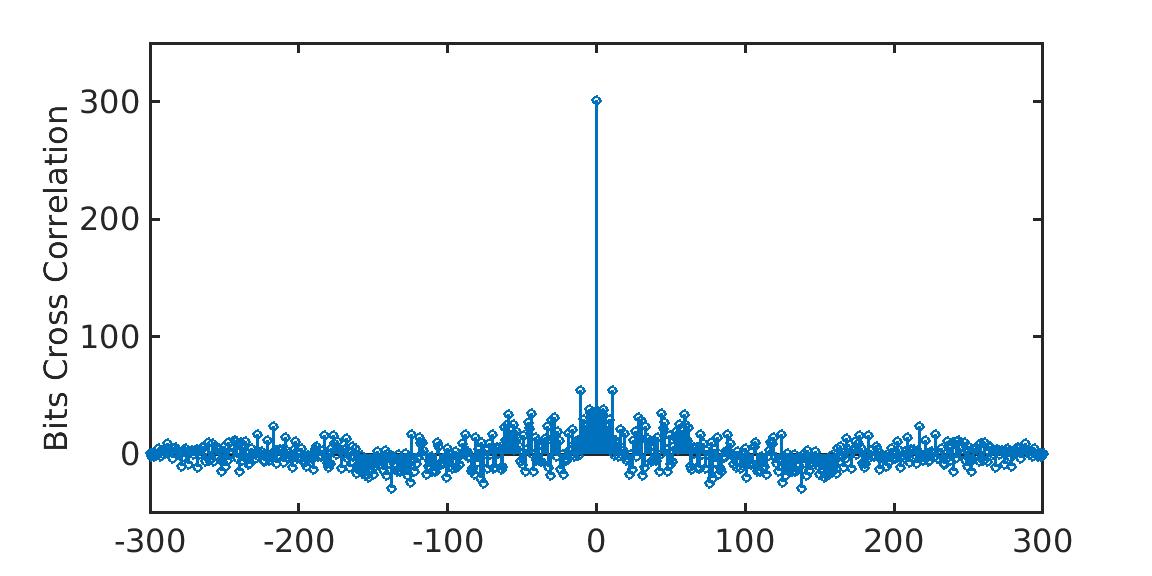}}
\caption{Correlation coefficient peak with a value of 300 showing that 300 bits of a single sub-frame received by attacker and by the victim are same.}
\label{bits_correlation}
\end{figure}

To validate our setup and to verify that the navigation bits stay untouched throughout our attack, we compare the bits received at the attacker and the bits received by the victim receiver using the cross-correlation function (Figure~\ref{bits_correlation}) for a single sub-frame (300 bits).
It is important to note that the signal generation methods that we used do not affect the feasibility of the attack on signals with message authentication since the navigation message contents remain untouched by the attacker.

\subsection{Evaluation Scenarios} 

Based on the setup we described above, we show that our attacker can generate spoofing signals for a target location far away from the victim (and the attacker)'s true location.
We evaluated our attacker setup for both static (stationary locations) and dynamic scenarios.

\begin{figure}[t]
\centerline{\includegraphics[width=3.1in]{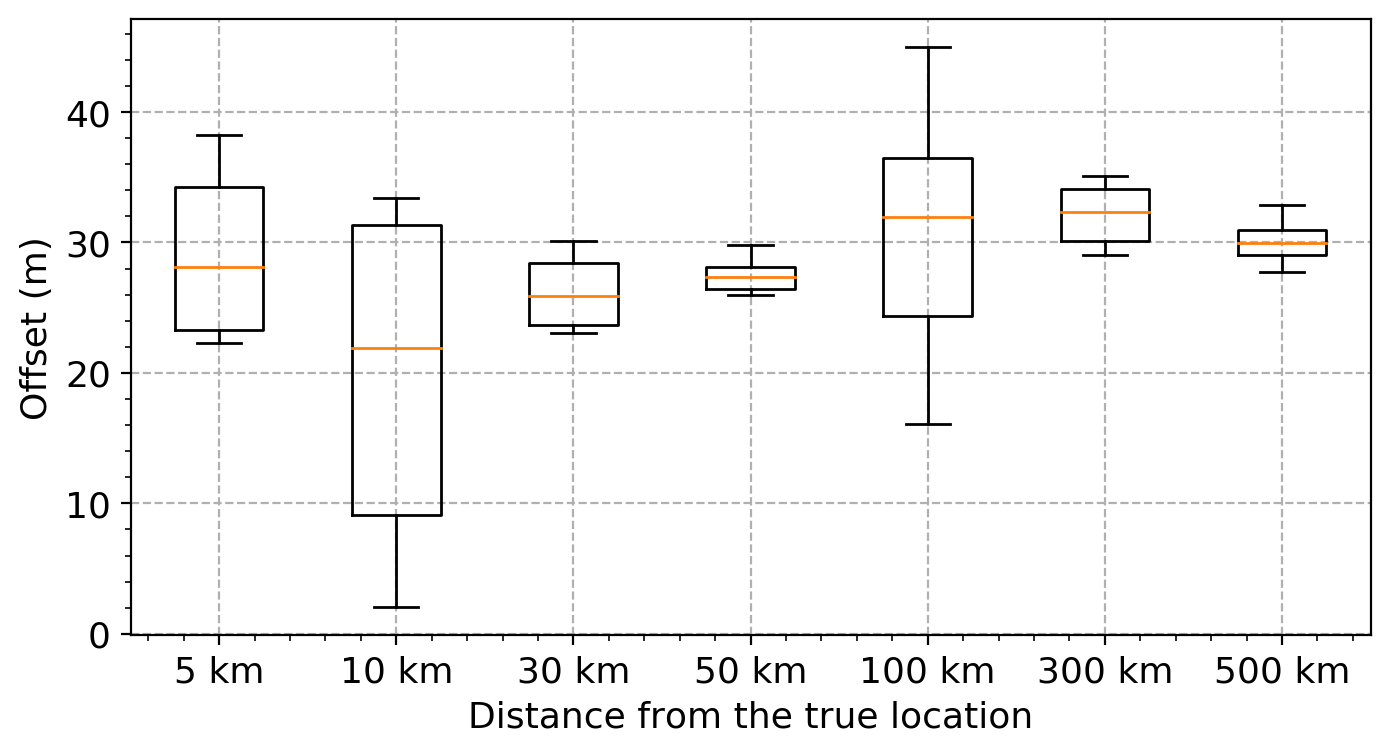}}
\caption{Accuracy of the spoofed location as a function of its distance from the attacker's location. The offset is the distance between the spoofed location and the actual location calculated by the victim receiver.}
\label{offset_acuracy_boxplot}
\end{figure}

\paragraph{Static scenarios.} For verifying the feasibility of our attack on static scenarios, we picked spoofing locations at various distances away from the receiver's true location. 
We evaluated the accuracy of the proposed attack by measuring the offset between the spoofing location and the obtained location at the victim receiver.
We were able to spoof the victim receivers (both ublox and GNSS-SDR) to our arbitrary locations, proving the attack's success. 
The results of our experiments are shown in Figure~\ref{offset_acuracy_boxplot}.
Specifically, we run the real-time experiments for each location for 10 minutes continuously to prove the stability of the obtained results, although the satellite's constellation keeps changing over time. We thoroughly examine the impact of satellite orbital motion in section~\ref{sec:eval_results}.
We once again emphasize that the contents of the navigation messages remained \emph{unchanged} throughout the attack. 

\paragraph{Dynamic motion scenario.}
We also evaluate the ability to generate spoofing signals that deceive the receiver into believing it is in motion at an arbitrary location away based on the legitimate signals received at the true location.
To spoof such a motion, the attacker has to manipulate the physical layer properties of the spoofed signal to reflect the updated position as per the desired trajectory.
As mentioned earlier, our implementation calculates pseudoranges, rate of change of pseudoranges, carrier phase, code phase delay and carrier frequency offsets to replicate the Doppler shifts for the specified position.
To enable the dynamic motion, we use a sequence of latitude and longitude values that reflect the target path such that target speed = $distance(p_{t}, p_{t+1})/dt$ where $p_{t}$ and $p_{t+1}$ are sequential positions of the trajectory as a real-time input to the signal synthesizer.
The signal-synthesizer then calculates the required parameters and modulates the incoming bits to generate the necessary spoofing signal.

\begin{figure}[t]
\centerline{\includegraphics[width=3.2in]{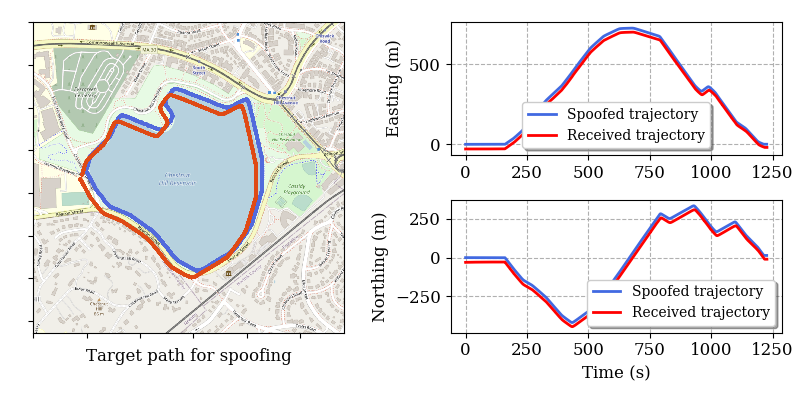}}
\caption{A comparison of spoofed trajectory and the locations estimates of the victim receiver}
\label{dynamic_scenario}
\end{figure}

To evaluate our strategy and its implementation, we generate and transmit a signal that forces the target into believing that it is moving at a speed of 1.98 m/s along a pre-determined path which is 2.43 km and in an area which is 5.5 km away from the victim's original location.
In this experiment, we only assume that the victim is within the radio range of the attacker and the victim can be either stationary or can be moving.
It is important to note that, the attacker doesn't need to have prior knowledge of the victim's true location for successful execution of this dynamic motion scenario.
Similar to the static location spoofing, we do not manipulate the legitimate navigation message contents and achieve dynamic motion by introducing appropriate temporal and Doppler changes to the legitimate signals.
Figure~\ref{dynamic_scenario} shows a comparison of the spoofed trajectory and the received location estimates

\paragraph{Proof-of-Concept Attack on Galileo Signals.} 
For testing the attack's success against Galileo, we generated signals corresponding to \emph{location anonymized for review} using NCS TITAN GNSS simulator~\cite{NCS_TITAN}.
Then, we configured the attacker to generate spoofing signals for a location 100 km away. 
With knowing the satellite ephemeris data, we calculated the delays in advance for the spoofed location.
The victim receivers were successfully spoofed to the desired location with an offset of $\approx$90 m. 
As we mentioned earlier, Galileo open service navigation message authentication (OSNMA) has started its test phase recently. 
But due to the limited availability of the OSNMA signals in space~\cite{nicola2022galileo}, we implemented our real-time setup based on GPS signals. 
Additionally, powerful GPS simulators provide the required flexibility to investigate the different aspects of the attack.

\subsection{Attack Performance Analysis}
\label{sec:eval_results}

In this section, we assess and evaluate the factors that impact the performance of the attack.
Our main evaluation metrics are the accuracy of the spoofed location estimated by the target and coverage.
We evaluate accuracy as the difference in the location spoofed by the attacker and the location estimated by the victim receiver.
Coverage is the furthest location an attacker can spoof the victim receiver \emph{from where the legitimate signals were received}.
The accuracy of the spoofed location, i.e., the difference between the spoofed target location and the location estimated by the victim receiver, depends on three main factors: i) attacker's sampling rate, ii) geometric dilution of precision (GDOP) of the spoofed satellites, and iii) satellite's orbital motion. 
We assume that the victim receiver and the adversary are in close proximity.

\begin{figure}[t]
\centerline{\includegraphics[width=3.0in]{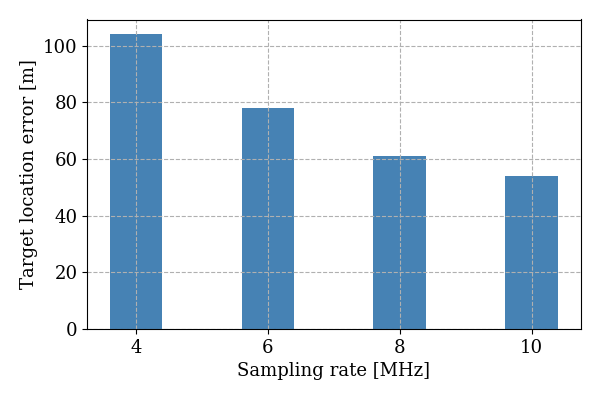}}
\caption{The accuracy of the spoofed location varies with the sampling rate. The sampling rate determines the resolution and accuracy of the pseudorange.}
\label{samp_rate}
\end{figure}

\paragraph{Impact of Attacker's Sampling Rate: }
The fundamental premise of the attack is to introduce specific delays to individual satellite signals. 
The delays are applied by temporally shifting the raw signal samples (i.e., IQ samples) appropriately.
Thus, a direct factor that influences the accuracy of the spoofed location is the ability of the attacker to precisely achieve the needed sample delay.
The adversary will achieve the required delays given a sufficient sampling rate.
We evaluate the effect of sampling rate on the accuracy and present our results in Figure~\ref{samp_rate}.
We observe that the accuracy of spoofed target location increases with the sampling rate.
For instance, given a sampling rate of 4 MHz, every sample delay corresponds to a 75 m change in pseudorange.
At 10 MHz sampling rate, the attacker can manipulate each pseudorange value with a resolution of 30 m. 
The results shown in Figure~\ref{samp_rate} are the final offsets from the target location as estimated by the victim receiver.

\begin{figure} [t]
\centerline{\includegraphics[width=2.8in]{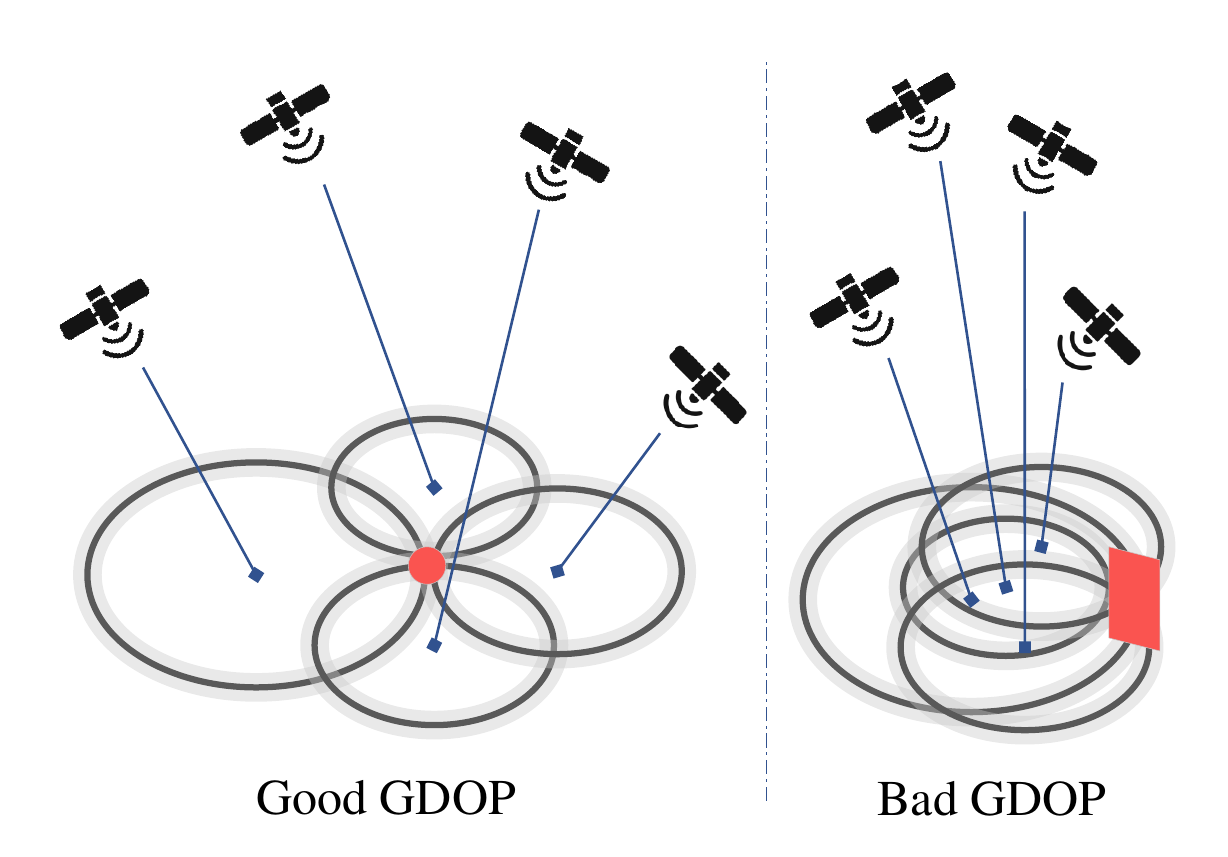}}
\caption{Examples of satellite constellations with good GDOP and bad GDOP. Geometrically more spread out satellite constellation causes lower dilution of precision than satellites grouped closer.}
\label{GDOP}
\end{figure}

\begin{figure*}[t]
\centering
\includegraphics[width=0.9\linewidth]{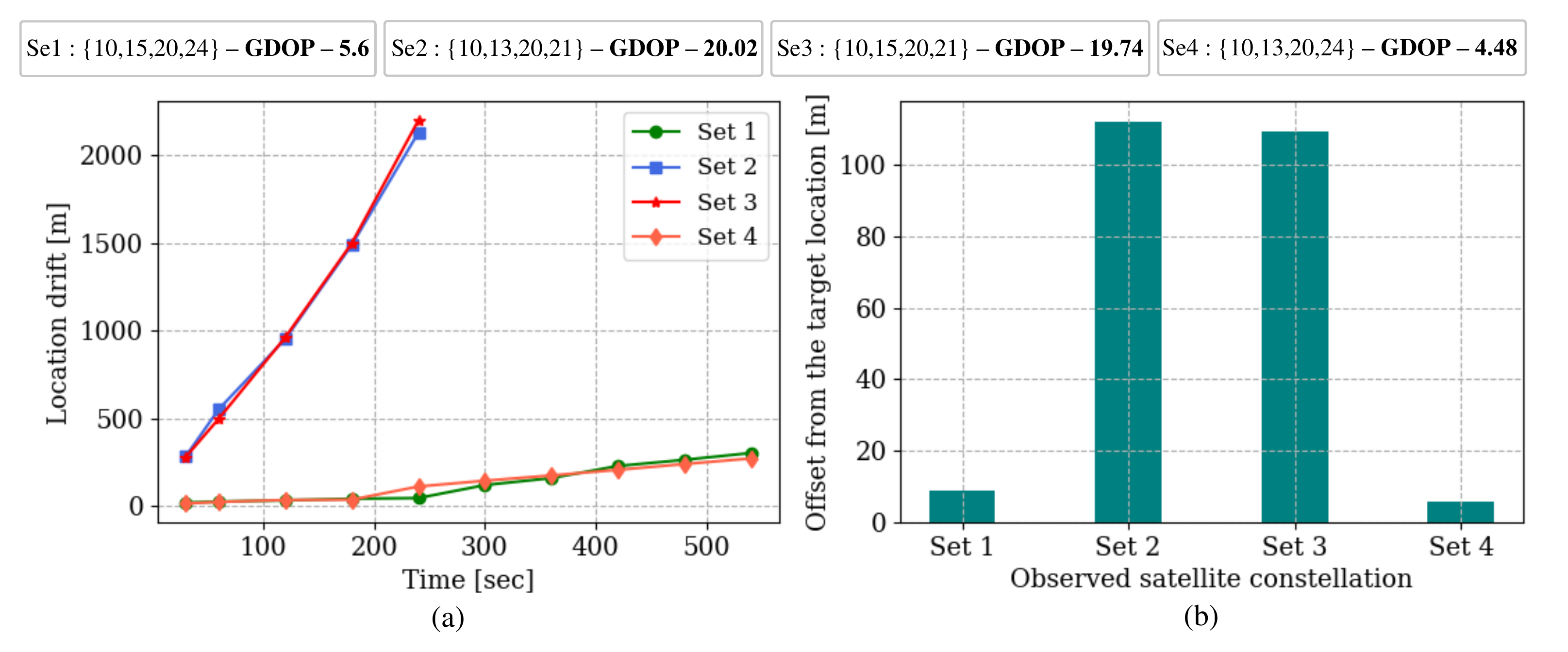}
\caption{The effect of satellite constellation geometry on the accuracy of spoofing. (a) Obtained location at victim receiver drifts over time. (b) Overall error in the spoofing location for four sets with varying GDOP.}
\label{combined}
\end{figure*}

\paragraph{Effect of GDOP: }
An important factor that directly affects the accuracy of the position estimates in any GNSS is the constellation of satellites' signals used to compute the location. 
The accuracy depends on the number of visible satellites and their elevations in the sky, i.e., the spread of satellites.
The GDOP is low for a satellite constellation spread apart and high for a constellation with satellites clustered in a single direction.
Figure~\ref{GDOP} shows an example of constellations with good and poor GDOP values.
The same principle applies to the attacker's spoofing signals. 
While choosing the satellite signals to manipulate temporally, it is essential to select satellite signals that have a low GDOP.
To determine the best constellation for a receiver, we calculate the GDOP of several constellations and choose a satellite constellation with the appropriate GDOP.
Figure~\ref{combined}~(b) shows the effect of GDOP on the spoofed location accuracy for four distinct sets of satellites with varying GDOP.
The results confirm our hypothesis that selecting the correct subset of satellites provides better control to the attacker regarding spoofing positioning accuracy.

\begin{figure}[t]
\centerline{\includegraphics[width=3.0in]{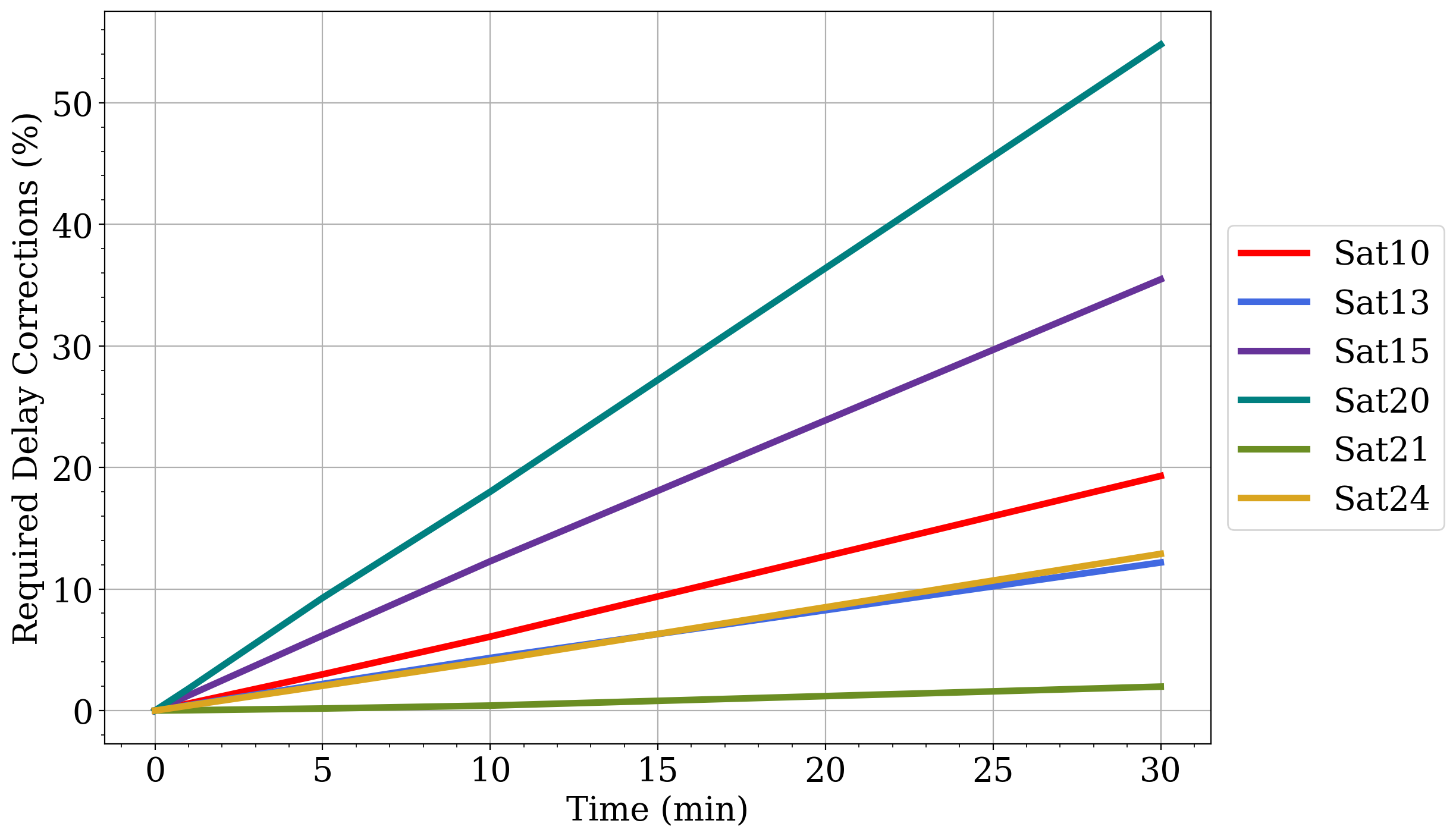}}
\caption{Delay corrections to maintain the spoofed location. The satellite geometry and orbit determine the required update rate for each satellite. Here, satellite 20 requires more aggressive corrections over time.}
\label{delay_rate}
\end{figure}

\paragraph{Impact of Satellite Orbital Motion: }
Recall that the pseudoranges are calculated based on the distance between the satellite and the receiver on the ground.
The satellite orbits are configured to have a certain number of satellites visible to any part of the earth. 
For example, GPS has its satellites orbit the earth along six orbital planes, and Galileo's satellites orbit the earth along three orbital planes.
As a result, the estimated pseudorange changes over time with a rate dependent on the location and time. 
Based on the satellite's velocity and position in the ephemeris data, the adversary can estimate the delay update rate and the required delay.
Figure~\ref{delay_rate} shows how the GPS pseudorange changes over 30 minutes at the \emph{location anonymized for review}.
Note the difference in the rate of change of pseudorange for each satellite.
Consequently, the attacker's choice of satellites directly impacts how often the adversary needs to update the delay calculation.
Figure~\ref{combined}~(a) shows the drift in the estimated location if the adversary does not recompute the delays, given a set of satellite signals spoofed.
For example, for satellite sets 1 and 4, the drift in the spoofed target location is less than 100 m even after 5 minutes. 
However, for satellite sets 2 and 3, the location drifts more than 500 m within the first minute.
The estimated location drifts faster when a high GDOP satellite set is spoofed.
High GDOP satellite sets are typically clustered together, resulting in significant pseudorange changes over time.
Thus, the choice of satellites plays a critical role in the performance of the attack. 

\begin{figure}[t]
\centerline{\includegraphics[width=2.3in]{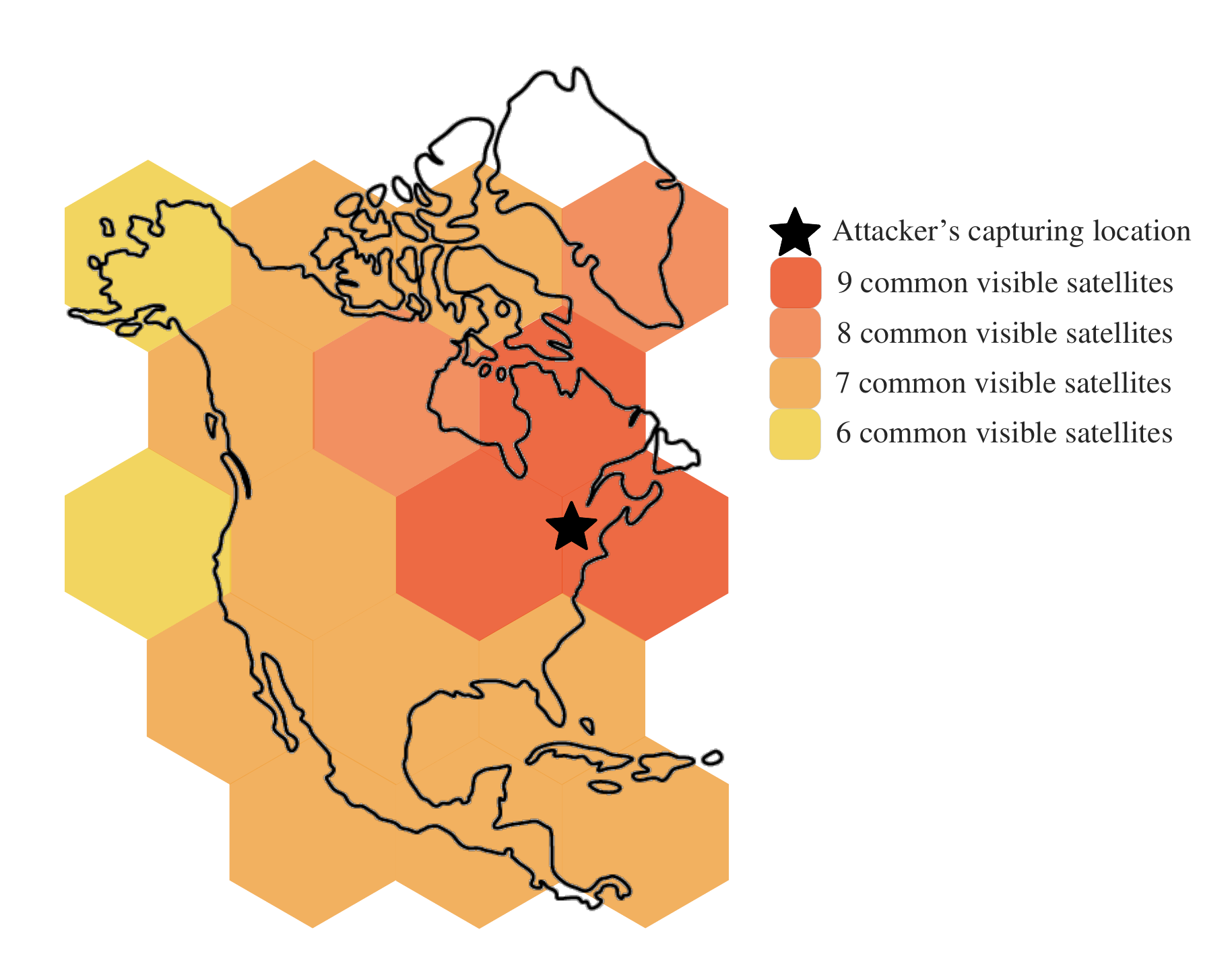}}
\caption{A comparison showing the number of common satellites between each region and the set of satellites visible at the attacker's location.}
\label{us_coverage}
\end{figure} 

\paragraph{Satellite Signal Coverage: }
In a typical GNSS spoofing attack, the attacker can spoof its target to any location on the earth as the attacker can generate navigation messages for any satellite. 
However, cryptographic signatures prevent the attacker from generating navigation messages, limiting its spoofing capabilities. 
Prior work~\cite{lenhart2021relay} showed the feasibility of relaying signals over long distances and spoofing the victim's location to where the signals were originally received. 
\noindent First, we evaluate the furthest distance an adversary can spoof a victim, assuming proximity between the attacker and the victim receiver. 
Given a set of satellites, the main factor that affects the ability of an attacker to spoof a specific location is the coverage of the set. 
In other words, given a victim receiver's actual location, an attacker can generate spoofing signals for any location provided there is a set of overlapping satellites also visible at the spoofed target location.
We analyzed the possible locations an attacker can spoof, assuming the true location is on the east coast of the United States. 
We divided North America into hexagons, each hexagon side measuring 1100 km. 
Then, we selected locations on the edges of these hexagons and determined the overlapping satellites between each of these locations and our assumed true location.
Our results (Figure~\ref{us_coverage}) indicate at least 7-8 common satellites across North America at a given time.
We used the satellites received at the assumed true location and generated signals for one of the northwest cities of the United States.
We validated the coverage of the attack by successfully spoofing to location $\approx 4000~km$ away from the actual location.

\begin{figure}[t]
\centerline{\includegraphics[width=2.8in]{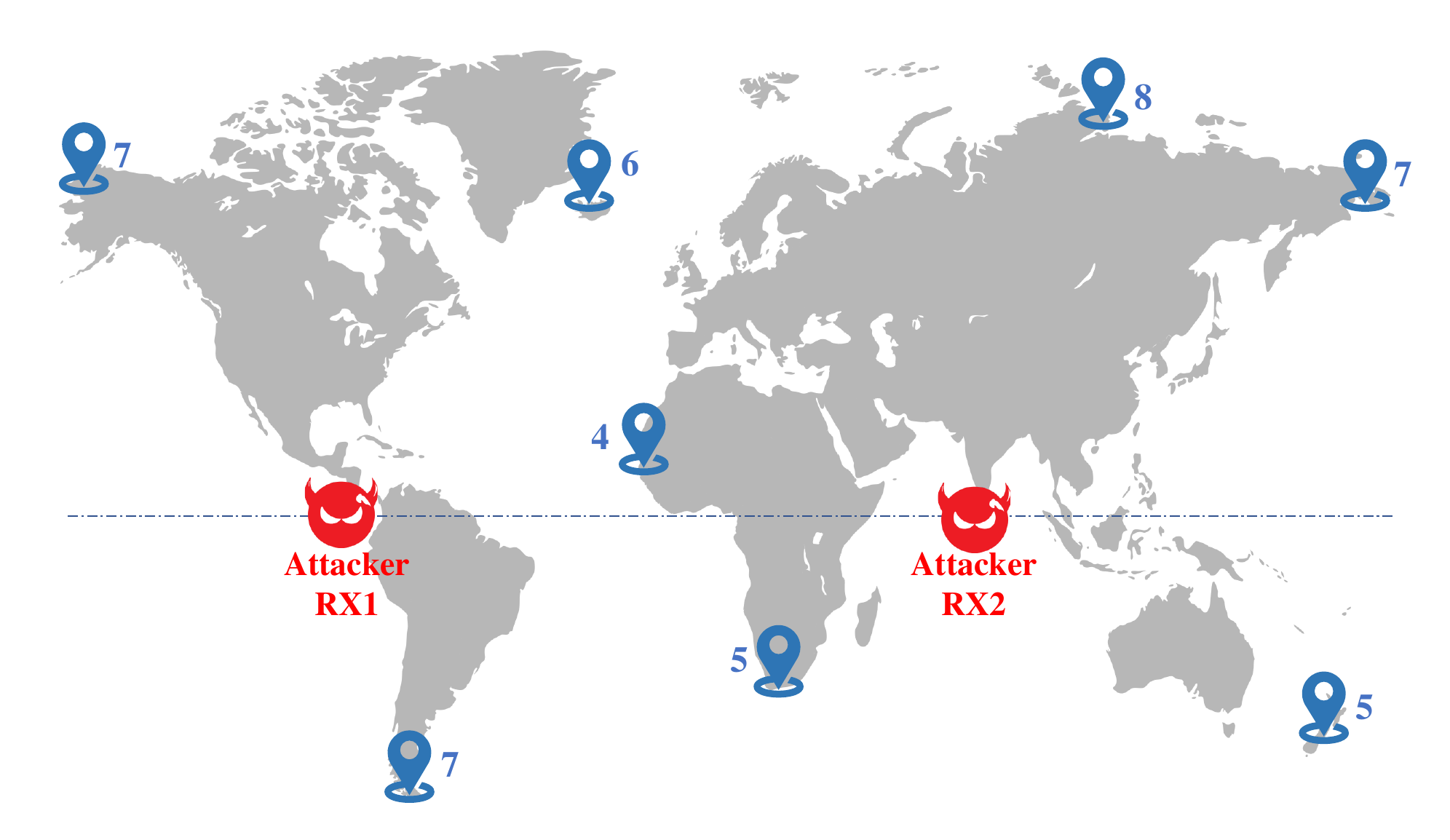}}
\caption{Potential positions for NAVMSG streamers that can provide worldwide coverage by providing raw navigation bits from at least 20 satellites at any time.}
\label{world_coverage}
\end{figure}

Furthermore, we analyzed the feasibility of generating spoofing signals to any arbitrary location in the world if the attacker could place multiple receivers anywhere in the world. 
Our analysis showed that with two receivers positioned along the equator, as shown in Figure~\ref{world_coverage}, we could observe at least four overlapping satellites at various corners of the world. 
This means that an attacker can generate spoofing signals to most of the earth's location with two receivers carefully placed and connected to the spoofer positioned close to the victim receiver.

\section{Countermeasures}
The direct protection of the time of arrival of the received signal (and the consequent protection of the pseudorange computation) cannot be completely ensured by using only cryptographic solutions. 
These solutions, alongside receiver-based techniques, can increase the resilience of GNSS receivers against spoofing attacks.
In this section, we discuss different types of countermeasures in GNSS receivers and their impact on detecting our proposed attacks.

\paragraph{Signal Power Monitoring:}
When multiple signals following a similar signal structure arrive simultaneously, a typical wireless receiver will automatically start tracking and demodulating the stronger signal. 
This phenomemon is refered to as the ``capture effect``.
GPS receivers are also affected similarly, as a result, they automatically start tracking the stronger signals.
In open sky conditions, satellite movement and ionosphere variations cause smooth changes in the received signal power~\cite{akos2012s}. 
Therefore, if a GNSS receiver experiences a sudden spike in the observed power, it can indicate the presence of the spoofing signal.
Often, in a spoofing attack, the attacker’s signals are strong enough to bury the legitimate signals under the noise floor, preventing the receiver from tracking them. 
Thus a receiver design will detect such an attack when a signal with a higher power is suddenly injected by the attacker. 
However, we can deter detection by incrementally changing transmit power of the attack signal.

\paragraph{Angle-of-Arrival Monitoring:}In most attack scenarios, an attacker is constrained by the number of attack nodes and their locations. 
The attacker uses a single (or fewer) attacker node to transmit counterfeit signals for the different satellites. 
In authentic settings, signals are transmitted from different satellites originating from different directions. 
Therefore, monitoring AoA can be employed to estimate the spatial signature of received signals and distinguish spatially correlated signals.
In~\cite{mcdowell2007gps} an antenna array structure is proposed to detect and mitigate spoofing signals based on their spatial correlation.
However, more research is needed to understand the possibility of using multiple antennas to create a false spatial signature and thwart such detection techniques.  

\paragraph{Checking with Other Navigation and Positioning Technologies:}
Auxiliary devices such as inertial measurement unit (IMU) can help the target receiver to discriminate against spoofing attacks. Additionally, the GNSS receiver can compare the solution to the other position and navigation solutions obtained by mobile networks or WiFi stations. 
Therefore, based on the confidence region of the solutions, there is a high chance of detecting the attack. It should be noted that these spoofing detection techniques increase the hardware and software complexity of receivers. The IMU sensors require calibration before being used for positioning purposes. In addition, alternative wireless location technologies, such as cellular networks, do not usually provide position solutions as accurate as GNSS solutions. Therefore, they might not be very helpful if there is a small mismatch between the fake and authentic position. Another issue is the limited coverage of cellular networks, which does not guarantee auxiliary positioning.

\paragraph{Auxiliary peak tracking: }
In auxiliary peak tracking ~\cite{ranganathan2016spree}, the receiver leverages the presence of authentic signals in addition to the attacker’s signals to detect spoofing attacks.
By overshadowing adversarial signals, an attacker can bury the legitimate signal under the noise in order to remove auxiliary peaks, so that the attack can remain undetected. 
Moreover, there are countermeasures that use the successive interference cancellation technique to mitigate GPS spoofing attacks ~\cite{sathaye2022semperfi}. 
This solution is specifically built for UAVs to recover from the GPS spoofing attacks. 
However, it is limited to a low-power overshadow attack.

\paragraph{Secret spreading codes: }
Researchers have proposed using partially unknown spreading codes, i.e., spreading codes with hidden markers~\cite{poltronieri2018analysis} that are disclosed in the future.
Techniques like these enable the receiver to authenticate the received signals based on the unknown bits of the spreading code that are revealed later.
Such a cryptographic countermeasure can limit the capabilities of an attacker as the attacker is unaware of the exact spreading code used by the legitimate signals.
This makes it difficult for our attacker to separate and regenerate satellite signals.
However, it is possible to implement codeless tracking techniques \cite{borio2011squaring} that can enumerate codes at run-time without waiting for the satellite to reveal the hidden markers or the unknown part of the spreading code.
Furthermore, directional antennas and spatial multiplexing techniques~\cite{van2020gnss, zhang2018efficient} can be also be used to separate satellite signals at the RF level.
An attacker can then add temporal shifts to these separated satellite signals and re-transmit them without spreading and modulation.
The one-way ranging implemented in GNSS systems proves to be a fundamental flaw in the design that makes the system vulnerable to various signal spoofing attacks.
As shown in~\cite{tippenhauer2012uwb}, with secure design and implementation, a bidirectional ranging system can be use for secure localization that has the potential to prevent distance manipulation through signal spoofing.

\section{Related Work}
Contrary to popular belief, cryptographic solutions are not always enough to safeguard a system.
Especially, satellite navigation systems that rely on one-way communications.
Several researchers in the past have demonstrated the ineffectiveness of cryptographic solutions.
The works that comes closest to our proposed attack are~\cite{humphreys2013detection,lenhart2021relay}. 
\cite{lenhart2021relay} demonstrated the feasibility of the relay attack on large distances.

In~\cite{humphreys2013detection}, the authors introduced the spreading code estimation replay (SCER) attacks where they statistically estimate the current bit at the time of transmission based on the previously received samples.
However, the proposed attacks have limitations in terms of locations that the attacker can spoof.
Moreover, this attack also has limitations in terms of the complexity of estimating parameters like chip length and power level.
This work was followed by~\cite{caparra2014improving},where the authors worked on system parameters to have a more effective SCER attack. 
They also proposed a countermeasure against SCER attack based on the assumption that the deployed strategy by the attacker is known to the victim receiver.

Authors in~\cite{papadimitratos2008protection} describe the threat landscape for cryptographically secure and unsecure GNSS signal.
They also provide basic countermeasure to prevent such attacks.
In \cite{zhang2019effects} the authors demonstrated that the GNSS authentication techniques are vulnerable to distance decreasing attacks.
Their proposed attack is focused on misleading the secure ranging and distance-bounding. 
Works like~\cite{fernandez2016galileo,shang2020flexible} discuss various approaches to execute replay attacks against cryptographically secured GNSS signal.
\cite{papadimitratos2008protection} studied the GNSS vulnerability to replay attacks. They evaluate the feasibility and effectiveness of replay/relay attacks against cryptographically secured GNSS.
Researchers have also explored various application layer attacks on GPS receiver software like~\cite{nighswander2012gps}, where the authors have used malicious GPS measurements and navigation message contents to exploit software bugs.

In \cite{shang2020flexible} the authors propose a delay control method that is capable of delaying signals by an integer multiple of the sampling period as well as a fractional multiple of the sampling period. Their worked was more focused on how to add the delays to the individual signals. 

In \cite{caparra2017feasibility} the authors focused on studying the self-spoofing attacks on GNSS Signals with Message Authentication. 
In self-spoofing, the GNSS receiving equipment is under the control of the adversary.
They showed that the symbol unpredictability provided by the cryptographic functions in NMA does not offer range assurance.
There has been limited work on detecting selective delay attacks.
Most notably~\cite{gallardo2020scer}, where the authors demonstrate an approach based on machine learning to detect a SCER attack using a set of features extracted from receiver search phase.
Several other works like~\cite{ranganathan2016spree, montgomery2011receiver, jansen2018crowd} describe the use of physical layer characteristics, multiple antennas and crowd-sourced networks to provide spoofing attack detection.

\section{Conclusion}
In this work, we designed and developed an attack that allows spoofing a victim receiver's location or motion without modifying the legitimate signal's navigation message contents. 
Specifically, we demonstrated how an attacker can temporally manipulate legitimate satellite signals received at a victim's true location in real-time to generate signals that correspond to arbitrary locations and motions far away from the victim's actual position. 
This is in contrast to prior work that required an attacker to be present and record legitimate satellite signals at the location they intend to spoof the victim's receiver. Particularly, our evaluations of the attack on both a commercial uBlox GNSS receiver and an open-source software-defined GNSS receiver (GNSS-SDR) show that it is indeed possible to spoof a victim receiver to locations more than 4000 km away from the true location without the need to modify the legitimate message contents or require high-speed communication networks. 

We also demonstrated the ability to generate spoofing signals that correspond to any arbitrary dynamic motion independent of the attacker or victim receiver's motion.
Finally, we analyzed the effect of factors like sampling rate, satellite constellation, and orbits on accuracy of the spoofed location and discussed the effectiveness of existing spoofing detection and mitigation techniques countermeasures against the proposed attack.

{\footnotesize \bibliographystyle{acm}
\bibliography{bibefile.bib}}


\end{document}